\documentclass[aps,prb,showpacs,twocolumn,longbibliography]{revtex4-1}
\usepackage[utf8]{inputenc}
\usepackage[colorlinks,pdfusetitle,urlcolor=blue,citecolor=blue,linkcolor=red,bookmarksnumbered,plainpages=false]{hyperref}
\usepackage{amsfonts}
\usepackage{amssymb}
\usepackage{amsmath}
\usepackage{graphicx}
\usepackage{epsfig}
\usepackage{subfigure}
\usepackage{appendix}

\begin{document}

\title{Resonant-amplified and invisible Bragg scattering based on spin
coalescing modes }
\author{K. L. Zhang}
\author{Z. Song}
\email{songtc@nankai.edu.cn}
\affiliation{School of Physics, Nankai University, Tianjin 300071, China}
\date{\today}

\begin{abstract}
Unlike a real magnetic field, which separates the energy levels of particle
with opposite spin polarization, a complex field can lead to a special kind
of spectral degeneracy, known as exceptional point (EP), at which two spin
eigenmodes coalesce. It allows an EP impurity to be an invisible scattering
center for a fermion with the resonant spin polarization, but an amplifying
emitter for opposite polarization. We show that a pair of conjugate EP modes
supports resonant mutual stimulation, acting as a resonant amplifier based
on the underlying mechanism of positive-feedback loop. Together with other
Hermitian eigenmodes, a fermion with EP polarization exhibits some exclusive
dynamics, referred to as EP dynamics. We construct several typical
superlattices, which are built up by embedding EP-impurity arrays in a
Hermitian two-dimensional square lattice. Numerical simulations are
performed to demonstrate resonant amplification and invisibility of Bragg
scattering.
\end{abstract}

\maketitle

\section{Introduction}

\label{Introduction} Spin as an intrinsic degree of freedom of a particle,
is the origin of magnetism and is a fundamental property of particles, which
is employed in spintronics \cite{vzutic2004spintronics} for information
processing. Carrying information in both the spin and charge of a particle
potentially offers devices with a greater diversity of functionality. In a
real magnetic field two energy levels of a half spin is separated,
associating with the spin can either be aligned along the field, or against
the field. Then a particle with arbitrary spin polarization is unavoidably
scattered by a magnetic impurity defined as Zeeman term $\mathbf{B}\cdot 
\mathbf{s}$, where $\mathbf{s}$ is a spin operator at local magnetic field $%
\mathbf{B}$. However, in stark contrast, an imaginary component of $\mathbf{B%
}$ field can shrink the energy gap between two opposite spin states.
Remarkably, the gap can be tuned to vanish, resulting in the coalescence of two
spin states. It relates to an exclusive concept in a non-Hermitian system,
exceptional point (EP), which has no counterpart in a Hermitian system. We
refer to such an impurity as EP impurity. The EP in a non-Hermitian system
occurs when eigenstates coalesce \cite{bender2007making, moiseyev2011non,
krasnok2019anomalies}, and usually associates with the non-Hermitian phase
transition \cite{feng2013experimental, gupta2019parity}. In a parity-time ($%
\mathcal{PT}$) symmetric non-Hermitian coupled system, the $\mathcal{PT}$
symmetry of eigenstates spontaneously breaks at the EP \cite%
{guo2009observation,ruter2010observation,peng2014parity,feng2014single,hodaei2014parity,feng2017non,longhi2018parity,el2018non,miri2019exceptional,ozdemir2019parity,wu2019observation}%
, which determines the exact $\mathcal{PT}$-symmetric phase and the broken $%
\mathcal{PT}$-symmetric phase in this system.

The EP has many applications \cite%
{miri2019exceptional,klaiman2008visualization,doppler2016dynamically,xu2016topological,assawaworrarit2017robust,midya2018non,cao2019angular, yoshida2018non, yoshida2019symmetry, yoshida2019exceptional}%
, not limited to non-reciprocal energy transfer \cite{xu2016topological},
unidirectional lasing \cite{miao2016orbital,longhi2017unidirectional}, and
optical sensing \cite{chen2017exceptional,hodaei2017enhanced}. Moreover, the
EP is a bifurcation point of the energy levels. Near the EP, the eigen
frequency response to the perturbation exhibits a square-root dependence 
\cite{chen2017exceptional} and a cubic-root \cite{hodaei2017enhanced}
dependence, respectively. In this regard, the EPs are useful for sensing in
comparison with the diabolic points; this feature has been verified in
optics, cavity optomechanics, cavity spintronics, and circuit quantum
electrodynamics \cite%
{wiersig2014enhancing,am2016parameter,liu2016metrology,lau2018fundamental,zhang2019quantum,djorwe2019exceptional,lai2019observation,hokmabadi2019non,cao2019exceptional,zhang2019dispersive}%
. The sensing susceptibility is greatly enhanced near the EPs \cite%
{chen2019sensitivity}. In addition to this, the dynamics of the systems with
parameters far away from, near and at the EP, exhibit extremely different
behaviors \cite{longhi2015half, regensburger2012parity,
zhang2013self, longhi2015non, longhi2015robust,
wang2016wave,yang2018dynamical}. When the system $\mathbf{B}\cdot \mathbf{s}$ is
far from or near EP but with finite energy gap, the dynamics is a periodic
oscillation with associated Dirac probability oscillating in the period of
time inversely proportional to the gap. When system is at EP, the Dirac
probability may be constant or increase quadratically with time, i.e., the
EP dynamics of the spin strongly depends on the initial state. This raises
interesting questions regarding the non-Hermitian spintronics based on
particular EP-related dynamics of electrons. Importantly, a complex magnetic
field is no longer a component of a toy model, but has been investigated in
a practice perspective \cite{lee2014heralded}.

In this paper, we investigate the effect of EP impurity on a particle wave
propagating in the Hermitian lattice in which various EP impurity arrays are
embedded. The EP dynamics allows an EP impurity to be an invisible
scattering center for a particle with the resonant spin polarization, but an
amplifying emitter for opposite polarization. We show that a pair of
conjugate EP modes supports resonant mutual stimulation, acting as a
resonant amplifier based on the underlying mechanism of a positive-feedback
loop. The important points are that (i) one of the two conjugate EP modes is
the auxiliary mode to another, i.e., mutual auxiliary states and (ii) two
conjugate EP modes are orthogonal in the context of Dirac inner product.
Together with other Hermitian eigenmodes that can be represented as two
conjugate EP modes, a particle with EP polarization exhibits some exclusive
dynamics, referred to as EP dynamics. We construct several typical
superlattices, which are built up by embedding an EP-impurity array in a
Hermitian two-dimensional ($2$D) square lattice. Numerical simulations are
performed to demonstrate resonant amplification and invisibility of Bragg
scattering \cite{longhi2015half, regensburger2012parity,
longhi2015non, longhi2015robust}, which, in low-dimensional
spinless systems, has been observed in the $\mathcal{PT}$-symmetric Bragg
scatterers experimentally \cite{regensburger2012parity}, and is also
proposed to realize in a tight-binding lattice with an imaginary magnetic
flux \cite{longhi2015non, longhi2015robust}.

This paper is organized as follows. In Sec. \ref{Hamiltonian and coalescing
spin modes}, we introduce a non-Hermitian spin-$1/2$ fermionic model and
discuss the coalescing spin modes. In Sec. \ref{Resonant mutual stimulation}%
, we analyze the mechanism of positive-feedback loop through a exactly
solvable $1$D model. In Sec. \ref{Amplified Bragg scattering}, we
demonstrate the dynamics of resonant amplification and invisibility of Bragg
scattering in $2$D square lattices. Finally, we summarize our results in
Sec. \ref{Summary}.

\section{Hamiltonian and coalescing spin modes}

\label{Hamiltonian and coalescing spin modes}

We begin this section by introducing a non-interacting spin-$1/2$ fermionic
model with complex impurities as a non-Hermitian term of the
Hamiltonian. We will show that it supports intriguing dynamic behavior
since it is a concrete example of a class of non-Hermitian Hamiltonian
presented in the Appendix. The Hamiltonian on an arbitrary lattice can be
written as 
\begin{equation}
H=\sum_{j<l}\sum_{\sigma =\uparrow ,\downarrow }\kappa _{jl}c_{j,\sigma
}^{\dag }c_{l,\sigma }+\mathrm{H.c.}+\sum_{j}\mathbf{B}_{j}\cdot \mathbf{s}%
_{j},  \label{H spin}
\end{equation}%
where operator $c_{j,\sigma }^{\dag }$\ creates a fermion of spin $\sigma $
at site $j$, and $\mathbf{s}_{j}=\left( s_{j}^{x},s_{j}^{y},s_{j}^{z}\right) 
$\ is the spin-$1/2$ operator, which is defined by $s_{j}^{\alpha }=\frac{1}{%
2}c_{j}^{\dag }\tau ^{\alpha }c_{j}$, satisfying the Lie algebra commutation
relation 
\begin{equation}
\left[ s_{j}^{\alpha },s_{j}^{\beta }\right] =\sum_{\gamma =x,y,z}i\epsilon
^{\alpha \beta \gamma }s_{j}^{\gamma },
\end{equation}%
and $\tau ^{\alpha }$ ($\alpha =x,y,z$) are the Pauli matrices; $c_{j}^{\dag
}$ is defined as $c_{j}^{\dag }=\left( c_{j,\uparrow }^{\dag
},c_{j,\downarrow }^{\dag }\right) $; and $\epsilon ^{\alpha \beta \gamma }$
is the Levi-Civita symbol. Here $\kappa _{jl}$ is the hopping strength between
two sites $j$ and $l$, determining the geometry of the lattice; and $\mathbf{%
B}_{j}$ is the on-site complex magnetic field, inducing non-Hermitian impurity.
In this work, we only consider three types of sites with 
\begin{equation}
\mathbf{B}_{j}=\left\vert B_{j}\right\vert \left\{ 
\begin{array}{c}
0, \\ 
(1,0,i), \\ 
(1,0,-i),%
\end{array}%
\right.
\end{equation}%
respectively.

The crucial point is that there are two types of EPs for terms $\mathbf{B}%
_{j}\cdot \mathbf{s}_{j}$ with nonzero $\left\vert B_{j}\right\vert $. We
note that two types of EP-impurities possess two different coalescing spin
modes $\frac{1}{\sqrt{2}}\left( c_{j,\uparrow }^{\dag }\pm ic_{j,\downarrow
}^{\dag }\right) \left\vert \text{\textrm{Vac}}\right\rangle $, i.e., under
the basis $\left( c_{j,\uparrow }^{\dag }\left\vert \text{\textrm{Vac}}%
\right\rangle ,c_{j,\downarrow }^{\dag }\left\vert \text{\textrm{Vac}}%
\right\rangle \right) $, we have 
\begin{equation}
\frac{1}{2}\left\vert B_{j}\right\vert \left( 
\begin{array}{cc}
\pm i & 1 \\ 
1 & \mp i%
\end{array}%
\right) \left( 
\begin{array}{c}
1 \\ 
\mp i%
\end{array}%
\right) =0.
\end{equation}%
Interestingly, such two EP modes are mutually auxiliary modes to each other,
respectively, i.e.,%
\begin{equation}
\frac{1}{2}\left\vert B_{j}\right\vert \left( 
\begin{array}{cc}
\pm i & 1 \\ 
1 & \mp i%
\end{array}%
\right) \left( 
\begin{array}{c}
1 \\ 
\pm i%
\end{array}%
\right) =\pm i\left\vert B_{j}\right\vert \left( 
\begin{array}{c}
1 \\ 
\mp i%
\end{array}%
\right) .
\end{equation}%
In addition, two EP modes are orthogonal in the context of Dirac inner
product as two eigenmodes for term $\mathbf{B}_{j}\cdot \mathbf{s}_{j}$ with
zero $\left\vert B_{j}\right\vert $.

\begin{figure}[t]
\centering
\includegraphics[width=0.5\textwidth]{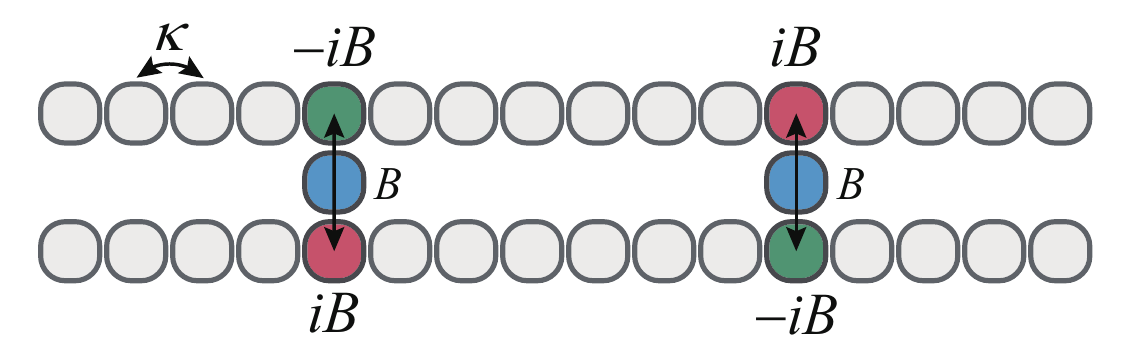}  
\caption{Schematic illustration of physical implementation for the
Hamiltonian in Eq. (\protect\ref{H spin}) by a resonator array. The red and
green dots represent pairs of resonators with $\mathcal{PT}$ imaginary
potentials $\pm iB$. The auxiliary resonators (blue dots) induce the
effective couplings $B$ between primary resonators.}
\label{fig1}
\end{figure}

Theoretically, a non-Hermitian Hamiltonian is the reduced
description for a selected sub-system of a Hermitian system, where the
complementary subspace is taken into account by means of an effective
interaction described by a non-Hermitian complex potential \cite%
{muga2004complex, jin2010physics, jin2011physical}. A negative imaginary
potential is an effective term in the Schr\"{o}dinger equation describing the
leakage of particle. In the case of that, the leakage can be controlled by
applying local real magnetic field on the channel to the environment, the
imaginary potential depends on its spin degree of freedom. Then one can
realize three kinds of spin-dependent negative imaginary potentials: (i) $%
-i3\gamma n_{j,\uparrow }-i\gamma n_{j,\downarrow }$, (ii) $-i2\gamma
n_{j,\uparrow }-i2\gamma n_{j,\downarrow }$, and (iii) $-i\gamma
n_{j,\uparrow }-i3\gamma n_{j,\downarrow }$. Here $n_{j,\sigma }=c_{j,\sigma
}^{\dag }c_{j,\sigma }$; and $\gamma$ is a real number. By subtracting a global term $-i2\gamma \left(n_{j,\uparrow }+n_{j,\downarrow }\right) $, the terms containing imaginary
magnetic field, $-i2\gamma s_{j}^{z}$ and $i2\gamma s_{j}^{z}$, in the
Hamiltonian can be obtained. Experimentally, the possible physical
implementation of the non-Hermitian model in this work can be coupled cavity
arrays \cite{longhi2009quantum, longhi2019topological, zeng2020topological,
ao2020topological}. As schematically illustrated in Fig. \ref{fig1}, the
spin degree of freedom is represented by the index of the chains, while the
imaginary magnetic field is realized by a pair of $\mathcal{PT}$ imaginary
potentials.

The general feature of such a system obeys the conclusion presented in the
Appendix, as an example wherein term $\mathbf{B}_{j}\cdot 
\mathbf{s}_{j}$ form a cluster with dimension $N_{\mathrm{cluster}}=2$.
Accordingly, there are two types of dynamics in a system with only one type
of EP-impurity: (i) When only the EP mode is involved in an initial state,
the Dirac probability is conservative. (ii) When only the auxiliary mode is
involved in an initial state, the Dirac probability is not conservative. In
this paper, we are interested in the case of two types of
EP-impurities embedded in the lattice, forming a special structure we dub
it EP superlattice. Importantly, the relation of mutually auxiliary modes
should result in mutual stimulation. We will demonstrate this point
analytically and numerically in the following sections.

\section{Resonant mutual stimulation}

\label{Resonant mutual stimulation}

In this section, we investigate the probability amplification by resonant
mutual stimulation based on a simple analysis for a system with two
different EP-sites embedded. An EP-site can create probability with the
auxiliary EP mode of another EP-site but remains the probability of an initial
state. As time goes on, it will result in a mutual stimulation. The total
probability should be amplified definitely when the system is finite due to
the reflection from the boundary.

\begin{figure}[t]
\centering
\includegraphics[width=0.5\textwidth]{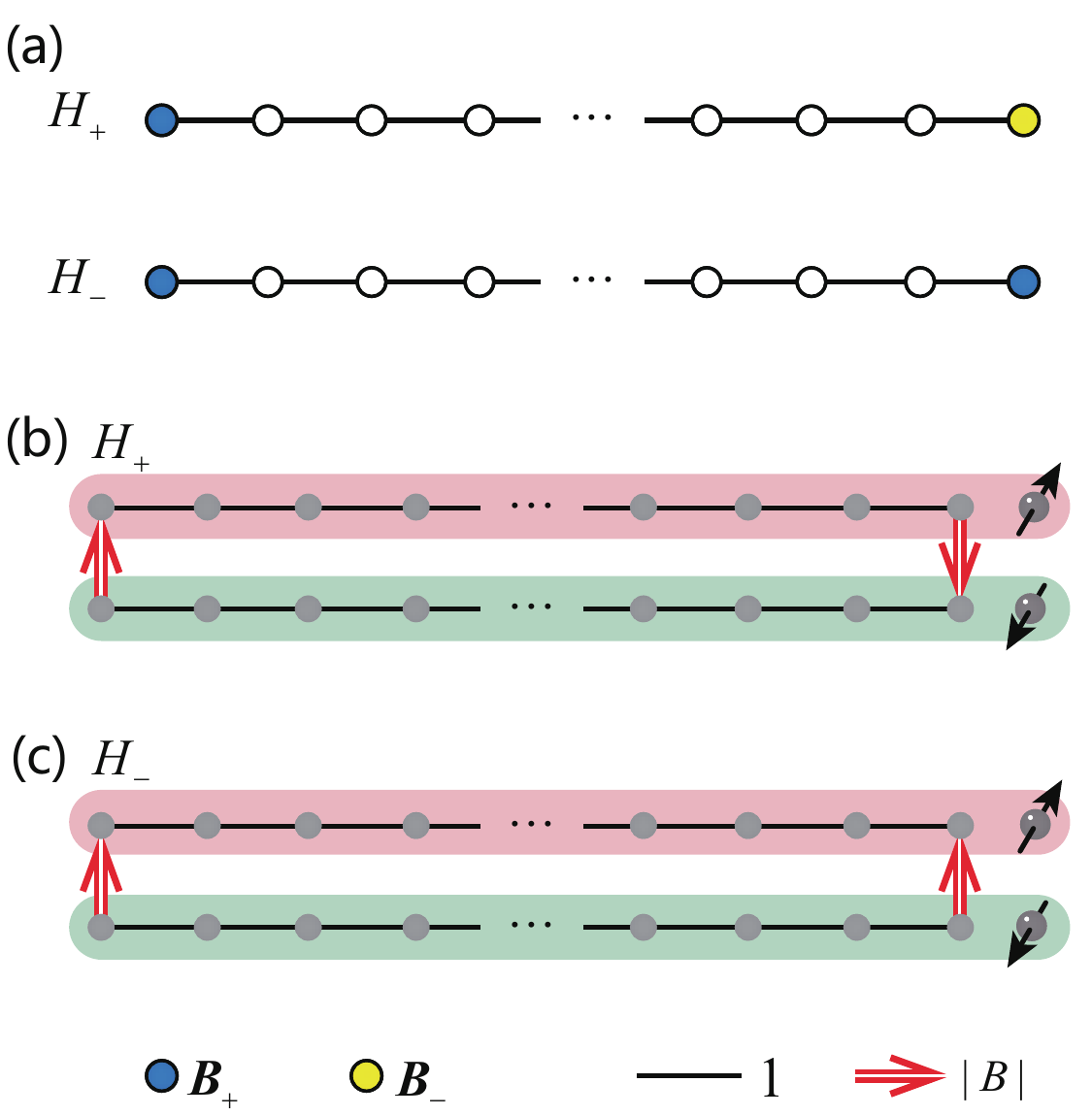}
\caption{Schematic illustrations for (a) Hamiltonian in Eq. (\protect\ref%
{Hpm}), where the EP-sites with complex fields $\mathbf{B}_{+}$ and $\mathbf{%
B}_{-}$ are represented by the blue and yellow dots, respectively. (b)
Equivalent Hamiltonians for $H_{+}$ and (c) $H_{-}$ in Eq. (\protect\ref%
{positive feedback loop}). Here the red arrow indicates that the systems
support unidirectional transition between two types of spin polarizations,
and $H_{+}$ act as a positive-feedback loop. }
\label{fig2}
\end{figure}

We demonstrate the concept quantitatively by an exactly solvable model,
which is an Hermitian chain with embedded two EP-sites at the ends. There are
two types of systems with the Hamiltonians

\begin{eqnarray}
H_{\pm } &=&H_{\mathrm{chain}}+\mathbf{B}_{+}\cdot \mathbf{s}_{1}+\mathbf{B}%
_{\mp }\cdot \mathbf{s}_{N},  \notag \\
H_{\mathrm{chain}} &=&\sum_{j=1}^{N-1}\sum_{\sigma =\uparrow ,\downarrow
}c_{j,\sigma }^{\dag }c_{j+1,\sigma }+\mathrm{H.c.}\text{,}  \label{Hpm}
\end{eqnarray}%
where the complex fields $\mathbf{B}_{\pm }=\left\vert B\right\vert (1,0,\pm
i)$. The two types of systems $H_{+}$ and $H_{-}$ are schematically
illustrated in Fig. \ref{fig2}(a). Applying the transformation%
\begin{equation}
d_{j,\lambda }=\sqrt{\frac{\lambda }{2}}\left( c_{j,\uparrow }+\lambda
ic_{j,\downarrow }\right) ,  \label{d_tran}
\end{equation}%
with $\lambda=\pm$, we have the equivalent Hamiltonians%
\begin{eqnarray}
H_{\pm } &=&H_{\mathrm{chain}}+\left\vert B\right\vert d_{1,+}^{\dag
}d_{1,-}+\left\vert B\right\vert d_{N,\mp }^{\dag }d_{N,\pm },  \notag \\
H_{\mathrm{chain}} &=&\sum_{j=1}^{N-1}\sum_{\lambda =+,-}\left( d_{j,\lambda
}^{\dag }d_{j+1,\lambda }+d_{j+1,\lambda }^{\dag }d_{j,\lambda }\right) 
\text{,}  \label{positive feedback loop}
\end{eqnarray}%
where operators $d_{j,\lambda }^{\dag }$ and $d_{j,\lambda }$ satisfy the
anti-commutation relations%
\begin{eqnarray}
\{d_{j,\lambda },d_{j^{\prime },\lambda ^{\prime }}^{\dag }\} &=&\delta
_{j,j^{\prime }}\delta _{\lambda ,\lambda ^{\prime }},  \notag \\
\left\{ d_{j,\lambda },d_{j^{\prime },\lambda ^{\prime }}\right\}
&=&\{d_{j,\lambda }^{\dag },d_{j^{\prime },\lambda ^{\prime }}^{\dag }\}=0.
\end{eqnarray}%
The transformation in Eq. (\ref{d_tran}) is unitary, which is different from
the similarity transformation in Eq. (\ref{can_pair}) that block
diagonalizes the original Hamiltonian (see Appendix \ref{Pseudo-Hermitian
modes}). Here, the biorthogonal conjugation operator $\bar{d}_{j,\lambda }$
defined in the Appendix reduces to $d_{j,\lambda }^{\dag }$ due to the fact
that two EP modes are orthogonal in the context of Dirac inner product. We
note that terms $\left\vert B\right\vert d_{1,+}^{\dag }d_{1,-}$ and $%
\left\vert B\right\vert d_{N,\mp }^{\dag }d_{N,\pm }$ represent
unidirectional transitions between different spin polarizations. Such a
transformation still holds for a general lattice with arbitrary geometry.
Going back to the present example, we find that Hamiltonian $H_{+}$ and $%
H_{-}$ are equivalent to two types of feedback loops (rings with two
unidirectional dimmers), which are schematically illustrated in Figs. \ref%
{fig2}(b) and \ref{fig2}(c), respectively. In particular, Hamiltonian $H_{+}$ forms a
positive-feedback loop that supports dynamics of mutual stimulation.

\begin{figure*}[tbh]
\centering
\includegraphics[width=1\textwidth]{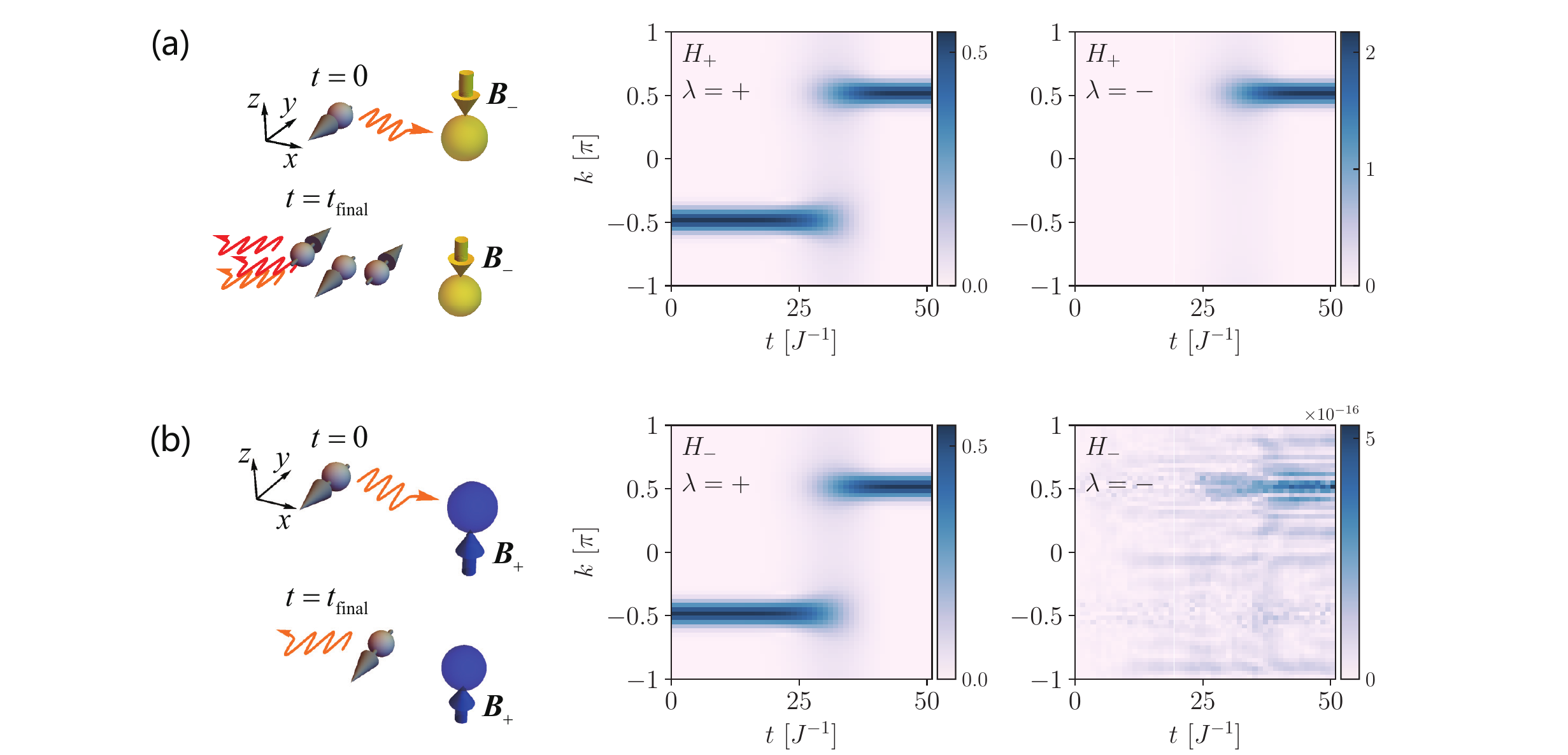}
\caption{Spin-orbit effect in the scattering process. The left panels of (a)
and (b) are the schematic illustrations of the scattering processes in
systems $H_{+}$ and $H_{-}$, respectively. The yellow (blue) arrow
represents local complex field $\mathbf{B}_{-}$ ($\mathbf{B}_{+}$).
Correspondingly, numerical result of the overlap $F_{\pm }^{\protect\lambda %
}(k,t)$ defined in Eq. (\protect\ref{F}) is presented in the right panels of
(a) $F_{+}^{\protect\lambda }(k,t) $ and (b) $F_{-}^{\protect\lambda}(k,t)$.
It indicates that after being scattered by the complex fields, the spin
polarization of initial excitation is conservative in both systems; and for
system $H_{+}$, the outgoing wave packet contains the other spin
polarization (which is zero for system $H_{-}$). The parameters of initial
excitation are $\protect\alpha=0.2$, $j^{\mathrm{c}}=30$ and $k^{\mathrm{c}%
}=-\protect\pi/2$; and are $\left\vert B\right\vert=1 $ and $N=60$ for the
both systems. Here the scale of the Hamiltonian is $J=2$.}
\label{fig3}
\end{figure*}

To understand the positive-feedback dynamics in such a system, we will
explore the features of the asymmetric dimer, which can be applied to the
original system in Eq. (\ref{positive feedback loop}). We focus on a
scattering problem for an asymmetry dimer which is embedded in an infinite
chain. For simplicity, we consider the Hamiltonian in single-particle
invariant subspace, which reads 
\begin{eqnarray}
H_{\mathrm{scatter}} &=&\sum_{j=1}^{\infty }\left( \left\vert j\right\rangle
\left\langle j+1\right\vert +\left\vert -j\right\rangle \left\langle
-j-1\right\vert \right) +\mathrm{H.c}.  \notag \\
&&+\mu \left\vert 1\right\rangle \left\langle -1\right\vert +\nu \left\vert
-1\right\rangle \left\langle 1\right\vert .
\end{eqnarray}%
Here position state $\left\vert \lambda j\right\rangle =d_{j,\lambda }^{\dag
}\left\vert \text{\textrm{Vac}}\right\rangle $ with $\left\vert \text{%
\textrm{Vac}}\right\rangle $ being vacuum state of operators $c_{j,\uparrow
} $ and $c_{j,\downarrow }$. Parameters ($\mu $, $\nu $) are asymmetric
hopping amplitudes, exhibiting the feature of asymmetric transmission in a
simple way. The Bethe ansatz scattering solution assumes the form 
\begin{equation}
\left\vert \psi _{k}\right\rangle =\sum_{j=1}^{\infty }[f_{k}\left( j\right)
\left\vert j\right\rangle +f_{k}\left( -j\right) \left\vert -j\right\rangle ]
\end{equation}%
where the wave function $f_{k}\left( j\right) $ reads 
\begin{equation}
f_{k}\left( j\right) =\left\{ 
\begin{array}{cc}
e^{ikj}+r_{k}e^{-ikj}, & j\leqslant -1 \\ 
t_{k}e^{ikj}, & j\geqslant 1%
\end{array}%
\right. .
\end{equation}%
Here $r_{k}$ and $t_{k}$ are the reflection and transmission amplitudes of
the incident wave with momentum $k$, which can be used to characterize the
property of the dimer. It is the solution for an incident wave from the left of
the system. Similarly, the solution for an incident wave from the right can be
obtained by exchanging $\nu$ and $\mu$. By solving the Schr\"{o}dinger
equation $H_{\mathrm{scatter}}\left\vert \psi _{k}\right\rangle
=E_{k}\left\vert \psi _{k}\right\rangle $, we obtain%
\begin{equation}
r_{k}=\frac{1-\mu \nu }{\mu \nu e^{2ik}-1}\text{, }t_{k}=\frac{2i\mu \sin k}{%
\mu \nu e^{2ik}-1}\text{.}
\end{equation}%
The result can be applied to the model in Eq. (\ref{positive feedback loop})
by taking $\left( \mu ,\nu \right) =\left( \left\vert B\right\vert ,0\right) 
$ or $\left( 0,\left\vert B\right\vert \right) $, representing the dynamics
for two types of initial state with spin polarizations $\lambda =\pm $,
respectively. (i) For $\left( \mu ,\nu \right) =\left( \left\vert
B\right\vert ,0\right) $, we have $\left\vert r_{k}\right\vert =1$ and $%
\left\vert t_{k}\right\vert =2\left\vert B\sin k\right\vert $, (ii) while $%
\left\vert r_{k}\right\vert =1 $ and $\left\vert t_{k}\right\vert =0$, for $%
\left( \mu ,\nu \right) =\left( 0,\left\vert B\right\vert \right) $. It
indicates that we always have $\left\vert r_{k}\right\vert =1$ in both cases.

Next, we further uncover and verify the above mechanism by considering the
process of a spin polarized Gaussian wave packet being scattered by the
right ends of chains $H_{+}$ and $H_{-}$, numerically. The initial
excitation is taken as $\left\vert \psi (0)\right\rangle =\left( \alpha /%
\sqrt{\pi }\right) ^{1/2}\sum_{j=1}^{N}e^{-\alpha ^{2}(j-j^{\mathrm{c}%
})^{2}/2}e^{ik^{\mathrm{c}}j}d_{j,+}^{\dag }\left\vert \text{\textrm{Vac}}%
\right\rangle ,$ where $j^{\mathrm{c}}$ is the center of wave packet; $k^{%
\mathrm{c}}$ is the central momentum; and $\alpha $ characterizes the width.
The site state $d_{j,+}^{\dag }\left\vert \text{\textrm{Vac}}\right\rangle $
is auxiliary mode of term $\mathbf{B}_{-}\cdot \mathbf{s}_{j}$ and EP mode
of term $\mathbf{B}_{+}\cdot \mathbf{s}_{j}$. The evolved state can be
represented in the form $\left\vert \psi _{\pm }(t)\right\rangle
=e^{-iH_{\pm }t}\left\vert \psi (0)\right\rangle $, which can be computed by
exact diagonalization numerically. To monitor the evolved state, we
introduce the overlap between evolved state and polarized state $\left\vert
k,\lambda \right\rangle =(1/\sqrt{N}) \sum_{j=1}^{N}e^{ikj}d_{j,\lambda
}^{\dag }\left\vert \text{\textrm{Vac}}\right\rangle $ in momentum space 
\begin{equation}
F_{\pm }^{\lambda }(k,t)=\left\vert \langle k,\lambda \left\vert \psi _{\pm
}(t)\right\rangle \right\vert,  \label{F}
\end{equation}%
where $k=2\pi m/N$ ($m$ is an integer in interval $(-N/2,N/2]$), and $%
\lambda =\pm$. Overlap $F_{\pm }^{\lambda }(k,t)$ characters the component
of evolved state $\left\vert \psi _{\pm }(t)\right\rangle $. That is, $%
F_{\pm }^{+}(k,t)$ and $F_{\pm }^{-}(k,t)$ count the same and different spin
polarizations as the initial excitation, respectively, as function of
momentum $k$ and time $t$. In other words, nonzero $F_{\pm }^{-}(k,t)$
reflects the emergence of spin-orbit effect. In the right panel of Fig. \ref%
{fig3}, we plot $F_{\pm }^{\lambda }(k,t)$ obtained from exact
diagonalization numerically. It indicates that the evolved state in system $%
H_{+}$ has evident spin-orbit effect [see Fig. \ref{fig3}(a)] and the evolved
state in system $H_{-}$ has no spin-orbit effect [see Fig. \ref{fig3}(b)].

\begin{figure*}[tbh]
	\centering
	\includegraphics[width=1\textwidth]{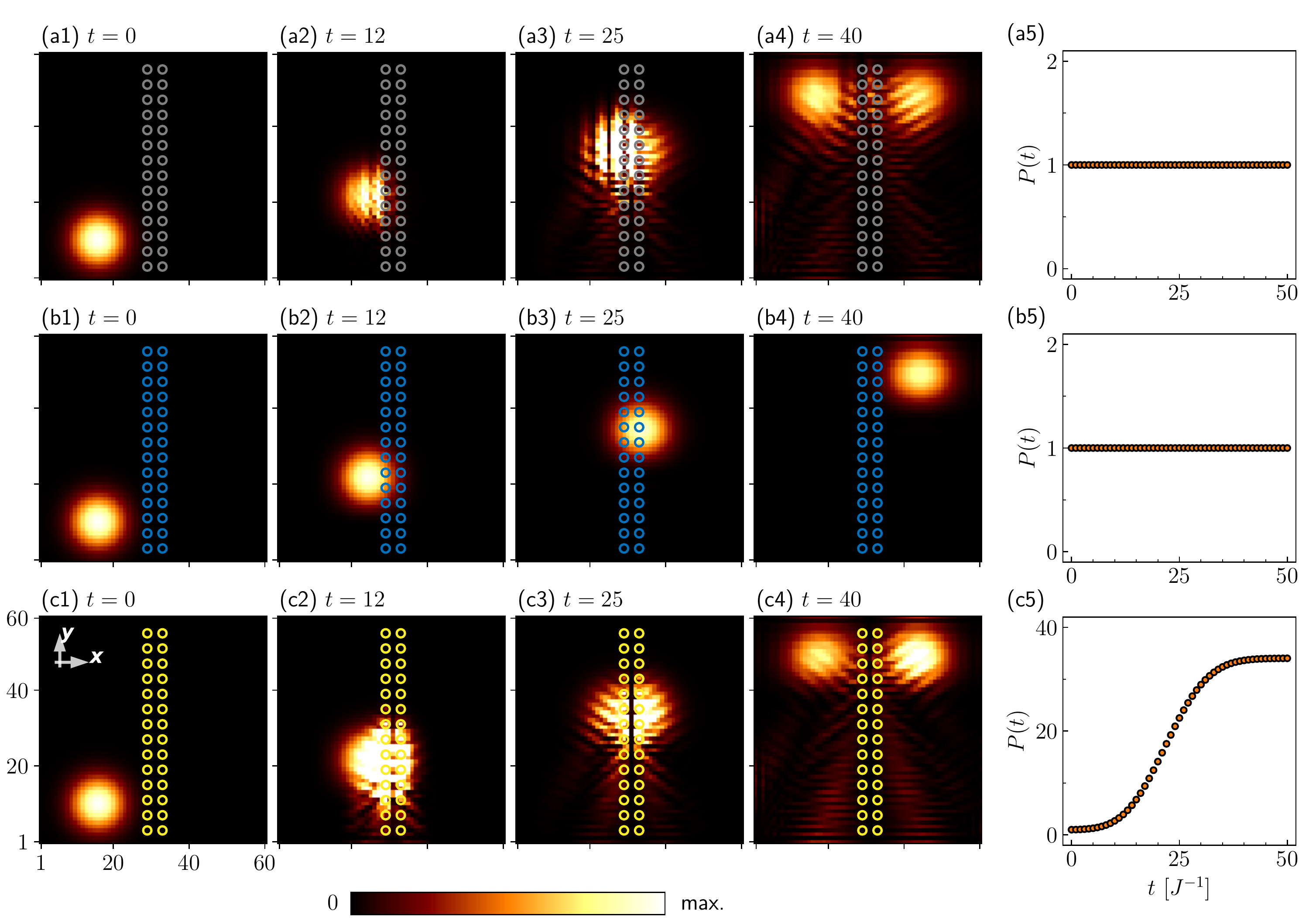}
	\caption{Bragg scattering in Hermitian $2$D system [(a1)-(a5)]. Invisible
		Bragg scattering [(b1)-(b5)], and amplified Bragg scattering [(c1)-(c5)] in
		non-Hermitian $2$D systems. The left panels are snapshots of intensity $%
		I_{j,l}\left( t\right) $ defined in Eq. (\protect\ref{intensity}) for the
		three systems, and the gray, blue and yellow circles represent real field $%
		\mathbf{B}$, complex fields $\mathbf{B}_{+}$ and $\mathbf{B}_{-}$,
		respectively. To better illustrate the Bragg scattering, we take different
		maximum value of the colorbars at different times. The right panels are the
		total Dirac probabilities $P(t)$ as functions of time. The parameters of the
		system are $\protect\kappa =1$, $\left\vert B\right\vert =5$ and $%
		N_{x}=N_{y}=60$, and for the initial excitation are $\protect\alpha =0.15$, $%
		(j^{\mathrm{c}},l^{\mathrm{c}})=(15,10)$, and $(k_{x}^{\mathrm{c}},k_{y}^{%
			\mathrm{c}})=(-\protect\pi /4,-\protect\sqrt{3}\protect\pi /4)$. The lattice
		constant of the Hermitian square lattice is $1$, and the distance between
		impurities is taken as $d=4$. Here the scale of the Hamiltonian is taken as $%
		J=2$. One can see that cases (a) and (c) have the same scattering pattern,
		except the amplification in the latter.}
	\label{fig4}
\end{figure*}

\begin{figure*}[tbh]
	\centering
	\includegraphics[width=1\textwidth]{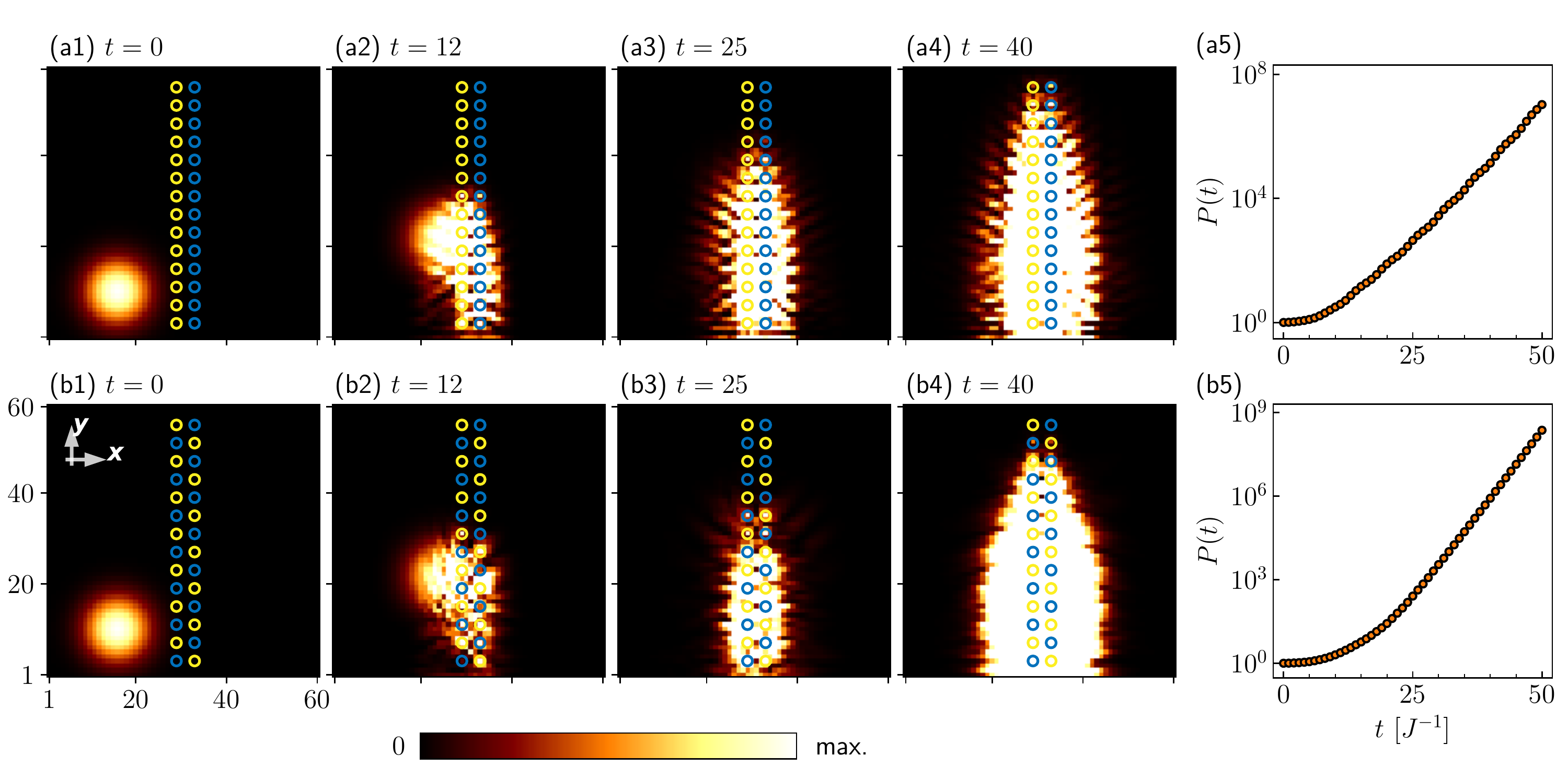}
	\caption{Resonant mutual stimulation in non-Hermitian $2$D systems in case
		(iii) [(a1)-(a5)] and case (iv) [(b1)-(b5)]. The left panels are snapshots
		of intensity $I_{j,l}\left( t\right) $ defined in Eq. (\protect\ref%
		{intensity}) for the two systems; and blue and yellow circles represent
		complex fields $\mathbf{B}_{+}$ and $\mathbf{B}_{-}$, respectively. The
		right panels are the total Dirac probabilities $P(t)$ as functions of time
		with logarithmic $y$-axis. Parameters for the both systems and the initial
		excitation are taken the same as that in Fig. \protect\ref{fig4}, except the
		configurations of EP-impurity fields. It indicates that the probability in
		(b) is one order higher than that in (a).}
	\label{fig5}
\end{figure*}

This provides a clear physical picture for understanding the feedback dynamics
of $H_{\pm}$. (i) For system $H_{+}$, any initial wave packet (except the
plane wave with $k=0$ and $\pi$) can trigger the Dirac probability explosion
since the EP modes at the ends of chain are mutual auxiliary modes. The
underlying mechanism is mutual stimulation. In addition, the evolved state
has evident spin-orbit effect. The momentum and spin polarization of a
particle are strongly correlated. (ii) For system $H_{-}$, the result
depends on the initial state: (a) For initial wave packet with EP mode, the
probability is conservative, since there is no mode switch. In contrast, the
evolved state has no spin-orbit effect. (b) For initial wave packet with
auxiliary EP mode, the probability with auxiliary EP mode is conservative,
while the probability with EP mode increases since there is always particle
with auxiliary EP mode that can provide the source to the probability with EP
mode. And over time, probability with EP mode is dominant. We would like
to point out that this amplification is fragile since it depends on the
conservation of probability with auxiliary EP mode. When a $2$D lattice is
considered, the probability with EP mode spreads and cannot supply the
switch to the probability with EP mode.

\section{Amplified Bragg scattering}

\label{Amplified Bragg scattering}

Now we extend our investigation to the 2D systems which support dynamics of
Bragg scattering. Here we consider a specific form of Hamiltonian Eq. (\ref%
{H spin}), wherein the EP-impurity array as a superlattice is embedded in a
Hermitian $N_{x}\times N_{y}$ square lattice with uniform nearest-neighbor
hopping $\kappa $. We will focus on four configurations of the superlattices
composed of two parallel EP-impurity arrays, in which, dynamics of invisible
Bragg scattering, amplified Bragg scattering, and resonant mutual
stimulation can be realized.

As a comparison and warm up, let us first consider the dynamics of Bragg
scattering in a Hermitian system with two parallel arrays formed by real
magnetic field $\mathbf{B}=(\left\vert B\right\vert ,0,0)$ [marked by gray
circles in Figs. \ref{fig4}(a1)-(a4)]. The initial excitation is an incident
wave packet with a fixed spin polarization 
\begin{eqnarray}
\left\vert \Psi (0)\right\rangle &=&\frac{1}{\sqrt{2\Omega }}%
\sum_{j=1}^{N_{x}}\sum_{l=1}^{N_{y}}e^{-\alpha ^{2}\left[ (j-j^{\mathrm{c}%
})^{2}+(l-l^{\mathrm{c}})^{2}\right] /2}e^{i(k_{x}^{\mathrm{c}}j+k_{y}^{%
\mathrm{c}}l)}  \notag \\
&&\times \left( c_{j,l,\uparrow }^{\dag }-ic_{j,l,\downarrow }^{\dag
}\right) \left\vert \text{\textrm{Vac}}\right\rangle ,
\end{eqnarray}%
where $\Omega =\sum_{j,l}e^{-\alpha ^{2}\left[ (j-j^{\mathrm{c}})^{2}+(l-l^{%
\mathrm{c}})^{2}\right] }$ is the Dirac normalization factor, $j$ and $l$
are site indexes of the lattice in $x$ and $y$ directions, $(j^{\mathrm{c}%
},l^{\mathrm{c}})$ is the center of wave packet, and $(k_{x}^{\mathrm{c}%
},k_{y}^{\mathrm{c}})$ is the central momentum. According to the Bragg's law 
\cite{bragg1913reflection}, the incident wave packet with certain direction $%
\theta =\arcsin \left[ n\pi /(\left\vert \mathbf{k}^{\mathrm{c}}\right\vert
d)\right] $ (angle between incident direction and $y$ axis) will induce
constructive interference in the direction of specular reflection. Here $n$
is a positive integer, $\left\vert \mathbf{k}^{\mathrm{c}}\right\vert =\sqrt{%
(k_{x}^{\mathrm{c}})^{2}+(k_{y}^{\mathrm{c}})^{2}}$, and $d$ is the distance
between two arrays. We set $n=1$, $d=4$, and $\left\vert \mathbf{k}^{\mathrm{%
c}}\right\vert =\pi /2$, and perform the numerical simulation of the time
evolution $\left\vert \Psi (t)\right\rangle =e^{-iHt}\left\vert \Psi
(0)\right\rangle $ by exact diagonalization. The dynamics can be observed
through the density distribution in real space defined as 
\begin{equation}
I_{j,l}\left( t\right) =\sum_{\sigma =\uparrow ,\downarrow }\left\vert
\left\langle \text{\textrm{Vac}}\right\vert c_{j,l,\sigma }\left\vert \Psi
(t)\right\rangle \right\vert ^{2},  \label{intensity}
\end{equation}%
as well as the total Dirac probability 
\begin{equation}
P(t)=\sum_{j=1}^{N_{x}}\sum_{l=1}^{N_{y}}I_{j,l}\left( t\right) .  \label{Pt}
\end{equation}%
The numerical results of $I_{j,l}\left( t\right) $ and $P(t)$ are presented
in Figs. \ref{fig4}(a1)-(a5). As expected, a "Bragg peak" arises in the
direction of specular reflection when time $t=40$, and of course, the total
Dirac probability is conservative.

Then we turn to the non-Hermitian systems. There are four configurations of
the superlattices composed of two parallel EP-impurity arrays: (i) both
arrays composed of field $\mathbf{B}_{+}$; (ii) both arrays composed of
field $\mathbf{B}_{-}$; (iii) one array composed of field $\mathbf{B}_{+}$
and the other composed of field $\mathbf{B}_{-}$; and (iv) staggered fields $%
\mathbf{B}_{\pm }$ form two arrays.\ Here the complex field is $\mathbf{B}%
_{\pm }=\left\vert B\right\vert (1,0,\pm i)$. The site state with fixed spin
polarization $\frac{1}{\sqrt{2}}\left( c_{j,l,\uparrow }^{\dag
}-ic_{j,l,\downarrow }^{\dag }\right) \left\vert \text{\textrm{Vac}}%
\right\rangle $ in the initial excitation $\left\vert \Psi (0)\right\rangle $
is auxiliary mode of term $\mathbf{B}_{-}\cdot \mathbf{s}_{j,l}$ and EP mode
of term $\mathbf{B}_{+}\cdot \mathbf{s}_{j,l}$. Taking the same system
parameters (except the impurity fields) and initial excitation as the above
Hermitian case, the numerical simulations of the time evolutions are
performed. Figs. \ref{fig4}(b1)-(b5) indicate that the superlattice in case
(i) is invisible for the evolved state, and the total Dirac probability is
conservative. Figs. \ref{fig4}(c1)-(c5) indicate that the evolved state in
system of case (ii) has the same scattering pattern as case (i), except the
amplification: the total Dirac probability as function of time is a
step-like function. These numerical results verify our prediction and are in
accord with the scattering solution of the $1$D system $H_{\mathrm{scatter}}$
in the previous section.

In view of the mechanism of positive-feedback loop, if the
system contains two kinds of EP-impurities at the same time, it is expected
to observe the dynamics of resonant mutual stimulation. Cases (iii) and
(iv) are for this consideration, and the results are plotted in Figs. \ref%
{fig5}(a1)-(a5) and Figs. \ref{fig5}(b1)-(b5), respectively. In both cases,
the dynamics of resonant mutual stimulation is evident, and the total Dirac
probability as a function of time is exponential after the wave packet being
scattered by the EP-impurity arrays. Clearly, the Dirac probability
increases more rapidly for the latter case, due to the sufficient mixing of
two kinds of EP-impurities.

\section{Summary}

\label{Summary}

In summary, we have developed a theory for a class of the non-Hermitian
Hamiltonian which supports a special dynamics due to the coexistence of
coalescing and zero modes. The most fascinating and important feature of
such systems is the resonant mutual stimulation. As an example, we have
investigated the dynamics of a non-Hermitian spin-$1/2$ fermionic
tight-binding model. It is shown that a pair of conjugate EP modes can act
as a resonant amplifier since its equivalent Hamiltonian is a
positive-feedback loop. The resonant tunneling between coalescing and zero
modes allows the construction of a superlattice, in which the non-Hermitian
EP-impurity array is embedded in a Hermitian lattice as a substrate. We
demonstrate the EP spintronics based on numerical simulations, including
resonant amplification and invisibility of Bragg scattering. Our findings
offer a platform for non-Hermitian spin devices.

\acknowledgments This work was supported by the National Natural Science
Foundation of China (under Grant No. 11874225).

\appendix

\section*{Appendix}

\label{Appendix} \setcounter{equation}{0} \renewcommand{\theequation}{A%
\arabic{equation}} \renewcommand{\thesubsection}{\arabic{subsection}}

In this Appendix, we consider a class of the non-Hermitian Hamiltonian which
supports a special dynamics. The model Hamiltonian in Eq. (\ref{H spin}) is one of the examples in this class. In general, the
Hamiltonian we are concerned with reads as follows 
\begin{eqnarray}
H &=&\sum_{j<l}H_{jl}+\sum_{j=1}^{N}H_{j},  \notag \\
H_{jl} &=&\kappa _{jl}\sum_{\alpha =1}^{N_{\mathrm{cluster}}}a_{j,\alpha
}^{\dag }a_{l,\alpha }+\mathrm{H.c.},  \notag \\
H_{j} &=&B_{j}\sum_{\alpha ,\beta =1}^{N_{\mathrm{cluster}}}J_{\alpha \beta
}a_{j,\alpha }^{\dag }a_{j,\beta },  \label{H}
\end{eqnarray}%
which consists of $N$ isomorphic clusters $H_{j}$, with each cluster having a
dimension $N_{\mathrm{cluster}}$. The label $j$ denotes the $j$th subgraph
of $N$ clusters, and $a_{j,\alpha }^{\dag }$\ ($a_{j,\alpha }$) is the boson
or fermion creation (annihilation) operator at the $\alpha $th site in the $j
$th cluster. The cluster $H_{j}$ is defined by the distribution of the
hopping integrals $\left\{ B_{j}J_{\alpha \beta }\right\} $ where the
parameter $B_{j}$\ is real, while\ the matrix $J_{\alpha \beta }$ is not
Hermitian. All $H_{j}$ have the same eigenstates and spectral structures due
to the fact that the geometry of each cluster is isomorphic. Meanwhile, terms 
$\sum_{j<l}H_{jl}$ is self-adjoint, i.e., $H_{jl}=H_{jl}^{\dag }$, which
describes the Hermitian connection between clusters. And such kind of
couplings are the type of similarity mapping, which is crucial for features
in the dynamics of the Hamiltonian $H$.

\subsection{Pseudo-Hermitian modes}

\label{Pseudo-Hermitian modes}

Firstly, we consider the case of $H_{j}$ being pseudo-Hermitian, i.e., $H_{j}
$ is non-Hermitian but can be diagonalized in the context of biorthonormal
inner product. Following the well established pseudo-Hermitian quantum
mechanics \cite%
{bender2007making,dorey2007ode,mostafazadeh2010pseudo,jin2013scaling}, we
always have 
\begin{equation}
H_{j}=B_{j}\sum_{\lambda =1}^{N_{\mathrm{cluster}}}\varepsilon _{\lambda }%
\bar{d}_{j,\lambda }d_{j,\lambda },  \label{cluster_diag}
\end{equation}%
with the operators $\bar{d}_{j,\lambda }$ and $d_{j,\lambda }$ being defined
as%
\begin{equation}
\bar{d}_{j,\lambda }=\sum_{\alpha =1}^{N_{\mathrm{cluster}}}h_{\alpha
\lambda }a_{j,\alpha }^{\dag },\text{ }d_{j,\lambda }=\sum_{\alpha =1}^{N_{%
\mathrm{cluster}}}g_{\alpha \lambda }^{\ast }a_{j,\alpha },  \label{can_pair}
\end{equation}%
where 
\begin{equation}
\sum_{\lambda =1}^{N_{\mathrm{cluster}}}g_{\alpha \lambda }^{\ast
}h_{\lambda \alpha ^{\prime }}=\delta _{\alpha \alpha ^{\prime }}.
\label{orth_re}
\end{equation}%
Here the existence of biorthogonal complete set of $\left\{ g_{\alpha
\lambda },h_{\alpha \lambda }\right\} $ rules out the exceptional point,
which will be discussed in the next section. Note that $\left\{ g_{\alpha
\lambda }\right\} $, $\left\{ h_{\alpha \lambda }\right\} $\ and $\left\{
\varepsilon _{\lambda }\right\} $\ are independent of the index $j$. Then
the operators $\bar{d}_{j,\lambda }$ and $d_{j,\lambda }$ are canonical
conjugate pairs, satisfying%
\begin{eqnarray}
\lbrack d_{j,\lambda },\bar{d}_{j^{\prime },\lambda ^{\prime }}]_{\pm }
&=&\delta _{jj^{\prime }}\delta _{\lambda \lambda ^{\prime }},  \notag
\label{can CR} \\
\lbrack d_{j,\lambda },d_{j^{\prime },\lambda ^{\prime }}]_{\pm } &=&[\bar{d}%
_{j,\lambda },\bar{d}_{j^{\prime },\lambda ^{\prime }}]_{\pm }=0,
\end{eqnarray}%
where $[$...$,$...$]_{\pm }$ denotes the commutator and anti-commutator,
depending on whether the excitation of the model is boson or fermion.

And accordingly, the original Hamiltonian can be rewritten as the form 
\begin{equation}
H=\sum_{\lambda =1}^{N_{\mathrm{cluster}}}H^{\lambda },
\end{equation}%
where the sub-Hamiltonian is 
\begin{equation}
H^{\lambda }=\varepsilon _{\lambda }\sum_{j=1}^{N}B_{j}\bar{d}_{j,\lambda
}d_{j,\lambda }+\sum_{j<l}\left( \kappa _{jl}\bar{d}_{j,\lambda
}d_{l,\lambda }+\kappa _{jl}^{\ast }\bar{d}_{l,\lambda }d_{j,\lambda
}\right) .
\end{equation}%
The relations%
\begin{equation}
\left[ H^{\lambda },H^{\lambda ^{\prime }}\right] =0,
\end{equation}%
indicates that the existence of invariant subspace indexed by $\lambda $.
When a given eigenvalue $\varepsilon _{\lambda }$\ is real, the dynamics in
the corresponding invariant subspace obeys the conservation of Dirac
probability \cite{jin2011hermitian}.

\subsection{Coalescing and zero modes}

\label{Coalescing and zero modes}

We then consider the system, which consists of two types of clusters
with zero eigenvalue: (i) being in EP and (ii) with $B_{j}=0$, respectively.
We denote the coalescing mode of cluster $j$\ at EP as $\left( \bar{d}_{j,%
\mathrm{C}},d_{j,\mathrm{C}}\right) $, i.e.,%
\begin{equation}
H_{j}=B_{j}\sum_{\lambda =1,\lambda \neq \mathrm{C}}^{N_{\mathrm{cluster}%
}-2}\varepsilon _{\lambda }\bar{d}_{j,\lambda }d_{j,\lambda }+0\times \bar{d}%
_{j,\mathrm{C}}d_{j,\mathrm{C}}.
\end{equation}%
Here, there are a total of $N_{\mathrm{cluster}}-1$\ modes, with an auxiliary
mode $\left( \bar{d}_{j,\mathrm{A}},d_{j,\mathrm{A}}\right) $\ being ruled
out. For a cluster with $B_{j}=0$, it has $N_{\mathrm{cluster}}$-fold
degenerate modes with zero eigenvalue. One of them can be $\left( \bar{d}_{j,%
\mathrm{C}},d_{j,\mathrm{C}}\right) $ or its auxiliary mode $\left( \bar{d}%
_{j,\mathrm{A}},d_{j,\mathrm{A}}\right) $. The total Hamiltonian $H$\ hence
possesses the common EP as $H_{j}$. When multi-non-Hermitian clusters are
involved, $H$\ supports high-order EP.

Now we consider the Hamiltonian in an invariant subspace spanned by a coalescing
mode, which has the form 
\begin{equation}
H^{\mathrm{C}}=\sum_{j<l}\left( \kappa _{jl}\bar{d}_{j,\mathrm{C}}d_{l,%
\mathrm{C}}+\kappa _{jl}^{\ast }\bar{d}_{l,\mathrm{C}}d_{j,\mathrm{C}%
}\right) .
\end{equation}%
It acts as a Hermitian Hamiltonian and exhibits Dirac probability preserving
dynamics. On the other hand, for the auxiliary mode, the inter-cluster
dynamics can still obey the Hamiltonian%
\begin{equation}
H^{\mathrm{A}}=\sum_{j<l}\left( \kappa _{jl}\bar{d}_{j,\mathrm{A}}d_{l,%
\mathrm{A}}+\kappa _{jl}^{\ast }\bar{d}_{l,\mathrm{A}}d_{j,\mathrm{A}%
}\right) .
\end{equation}%
However, the dynamics of an inner non-Hermitian cluster obeys the
non-Hermitian quantum mechanics, violating the conservation of Dirac
probability. In summary, if the initial state only involves the coalescing
mode, it will evolve in a Hermitian manner, while it blows up the probability for an 
auxiliary mode. In the main text, the example with $N_{\mathrm{cluster}}=2$
demonstrates this point in detail.

\end{document}